\documentclass[reprint,amsmath,amssymb,aip,apl]{revtex4-1}
\usepackage{graphicx}
\usepackage{dcolumn}
\usepackage{bm}

\begin{document}

\title{Fabrication of whispering gallery mode cavity using crystal growth}
\author{Hiroshi Kudo}
\author{Yohei Ogawa}
\author{Takumi Kato}
\affiliation{Department of Electronics and Electrical Engineering, Faculty of Science and Technology,\\
Keio University, 3-14-1 Hiyoshi Kohoku, Yokohama, 223-8522, Japan}
\author{Atsushi Yokoo}
\affiliation{NTT Nanophotonics Center, NTT Corporation, 3-1 Wakamiya Morinosato, Atsugi, 243-0198, Japan}
\affiliation{NTT Basic Research Laboratories, NTT Corporation, 3-1 Wakamiya Morinosato, Atsugi, 243-0198, Japan}
\author{Takasumi Tanabe} \email{takasumi@elec.keio.ac.jp}
\affiliation{Department of Electronics and Electrical Engineering, Faculty of Science and Technology,\\
Keio University, 3-14-1 Hiyoshi Kohoku, Yokohama, 223-8522, Japan}
\date{\today}

\begin{abstract}
We developed a new method for fabricating crystalline whispering gallery mode cavities based on laser-heated pedestal growth. We fabricated sapphire cavities and obtained a $Q$ factor of $1.6\times \mathrm{10^{4}}$ with a cavity whose diameter was about $240~\mathrm{\mu m}$. We showed numerically that the cross-sectional shape of the cavity is sensitive to the cavity $Q$, and we controlled it successfully by changing the growth condition in the molten zone, without significantly degrading the crystal structure.
\end{abstract}
\maketitle

Recent progress on micro- and nano-fabrication technologies has made it possible to utilize ultra-small cavities with an ultrahigh quality factor ($Q$) for applications such as low-power all-optical switches,\cite{ref20, tanabe_apl, nozaki} compact optical memories,\cite{ref21, tanabe_ol, yoshiki} and optical sensors.\cite{armani2007, ref10} Those applications are possible because a high $Q/V$ cavity exhibits various optical nonlinearities even at a very low input power ($V$ is the mode volume of the cavity). Among various type of microcavities, a whispering gallery mode (WGM) microcavity exhibits the highest $Q$.\cite{ref12} An ultra high $Q$ WGM cavity is particularly attractive as a building block for an ultra-narrow linewidth laser source\cite{liang2010} and a stable frequency comb source.\cite{ref22} A toroid microcavity\cite{ref1} and a bottle microcavity,\cite{ref2} both made of $\mathrm{SiO_{2}}$, exhibit record high $Q$s of about $\mathrm{10^{8}}$. As has been proven with optical fibers, the loss of $\mathrm{SiO_{2}}$ is ultralow; hence, it is an attractive material for fabricating WGM cavities. Although the $Q$ of a WGM cavity is currently limited by losses such as water absorption loss and fabrication errors, it has the potential to reach $Q \simeq 10^{10}$, which is the limitation caused by the material absorption of $\mathrm{SiO_2}$.

But crystalline materials such as $\mathrm{CaF_2}$ and $\mathrm{MgF_2}$ should exhibit even higher $Q$ value due to the ultra-low absorption coefficients of the material.\cite{ref16} They also have various unique characteristics such as, large mechanical stiffness and a large $\chi^{(2)}$ coefficient, which are attractive for opto-mechanics studies\cite{ref4} and electro-optic modulation.\cite{ref7, ref23} However, crystal resonators are not as simple to fabricate as $\mathrm{SiO_2}$ resonators, because we cannot use a conventional semiconductor process or a laser reflow process. The use of precise machining has been demonstrated,\cite{ref8} but the fabrication of smooth surfaces is extremely challenging for small WGM microcavities because ductile mode cutting is difficult with crystalline material that is usually fragile.\cite{kakinuma}

In this work, we develop a new technique for fabricating small crystalline WGM microcavities by employing laser-heated pedestal growth (LHPG).\cite{ref6, ref5} LHPG was originally developed for growing a thin rod-shaped single crystal for manufacturing fiber lasers, and can be applied to various materials such as $\mathrm{Al_{2}O_{3}}$, $\mathrm{LiNbO_3}$ and YAG.\cite{ref9} Since this method does not require complex precise machining it is easy and robust. If we apply LHPG to the fabrication of WGM microcavities, a smooth cavity surface will be obtained because it uses surface tension during crystal growth. However, crystals tend to grow faster in specific directions, and we need to find methods to control the longitudinal and cross sectional shape of the grown crystal to form a WGM cavity.

For our proof-of-principle demonstration, we used sapphire ($\mathrm{Al_2 O_3}$), which is a good material for constructing optical devices. It can be used as a host for an optical gain material, and moreover it exhibits an ultrahigh $Q$ at infrared wavelengths \cite{ref16}, which is an interesting wavelength regime for sensing applications.

\begin{figure}[htbp]
\centering
 \includegraphics[width=3.2in]{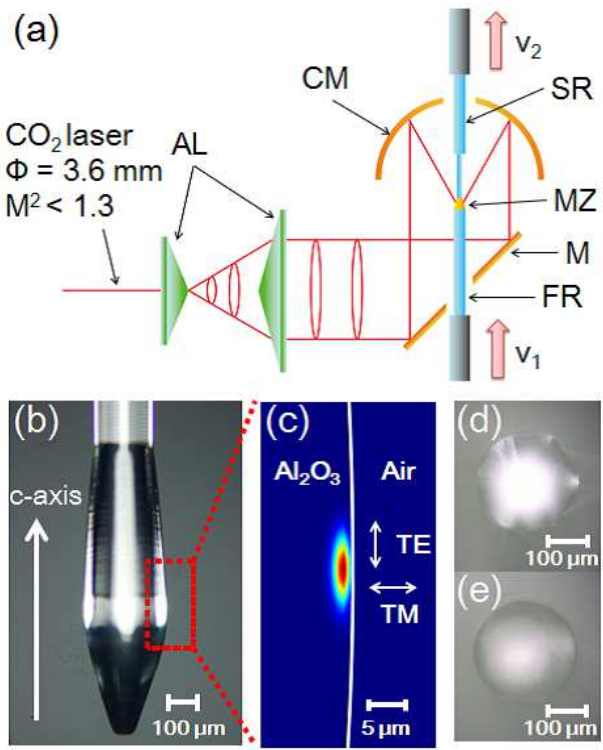}
 \caption{(Color online) (a) Schematic illustration of experimental setup. AL: ZnSe axicon lens, FR: feed rod, SR: seed rod, M: gold flat mirror with a hole at the center, CM: concave gold mirror (curvature = 100~mm) with a hole at the center, MZ: molten zone. A $\mathrm{ CO_{2}}$ laser with diameter $\phi$ and beam parameter product $M^2$ is used to heat the crystal rods. (b) Side view of the fabricated sapphire WGM microcavity. (c) TM-mode profile calculated with the finite-element method. (d) Cross section of the fabricated sapphire WGM microcavity without pre-heating the seed rod. (e) As (d) but the seed rod is pre-heated.}
\label{fig:1}
\end{figure}

Figure~\ref{fig:1}(a) is the LHPG setup that we used for our experiment. We used a $\mathrm{CO_2}$ laser (Coherent DIAMOND C-70) for heating the rods. The $\mathrm{CO_{2}}$ laser beam is formed into a donut shape by using a pair of axicon ZnSe lenses. This prevents the beam from hitting the base rod directly before it enters the setup. The donut shaped laser beam is focused on the top of the base rod with $\mathrm{360^{o}}$ axial symmetry by using a concave mirror with a curvature radius of 50~mm. This creates a spot with a size of $4.65 \times \mathrm{10^{-4}}~\mathrm{cm^{2}}$ in size, which enables us to obtain a power density of $8.58 \times 10^7~\mathrm{W/m^{2}}$ at the molten zone when the $\mathrm{CO_{2}}$ laser power is 4.0~W. The sapphire rods are fitted with a chuck to the computer controlled translation stages (Oriental Motor ASM46MA), and then placed near the focus of the concave mirror through a hole drilled at the center of the mirrors. The diameters of the base and seed sapphire rods are both $425~\mathrm{\mu m}$. Theoretically, the seed rod diameter does not affect to the diameter of the grown rod.

The tip of the feed crystal rod is heated by the $\mathrm{CO_{2}}$ laser beam (this forms a molten zone), and then a crystal fiber is grown by pulling the feed rod upwards at a speed of $v_1$, while the seed rod is moved in the same direction with a speed of $v_2$ ($v_1 > v_2$). The diameter $D$ of the grown fiber is given as, ${D_1}=({D_2}\sqrt{v_2})/(\sqrt{v_1})$, where ${D_2}$ is the diameter of the base rod. Our strategy is to modulate the speed of the feed and seed rods during growth, in order to fabricate a bulge, which can be used as a WGM cavity.

To demonstrate this idea, we start with standard LHPG. The laser power is set at 4~W and the velocities are set at $v_1=12~\mathrm{\mu m /s}$ and $v_2 = 2~\mathrm{\mu m /s}$. $v_2$ is set six times slower than $v_1$, and this results in the fabrication of a straight rod with a diameter of $174~\mathrm{\mu m}$.

Then we try to form a WGM cavity. For this purpose, we reduced $v_2$ to $6~\mathrm{\mu m/s}$ while keeping the speed of the base rod the same at $v_2 = 2~\mathrm{\mu m /s}$. This corresponds to a speed ratio of $v_1/v_2=1/3$ and will result in a rod with a diameter 1.4 times larger than before. Hence we are able to change the rod diameter locally, which requires us to localize the light in a longitudinal direction. By stopping the feed rod at the end, we obtained a bulge shaped WGM cavity as shown in Fig.~\ref{fig:1}~(b). The diameter of the fabricated cavity was about 0.24~mm, which is close to the value we designed. Figure~\ref{fig:1}~(c) shows the cross sectional intensity distribution of the WGM calculated by using the finite-element method (COMSOL Multiphysics 4.1). This result confirms that the light is localized in both the longitudinal and radius directions. The mode volume of this WGM is $1.28 \times 10^{-2}~\mathrm{mm^{3}}$.

Now we look closely at the cross section of the fabricated device. The cross sectional microscope images are shown in Fig.~\ref{fig:1}(d) and (e). The difference between the two is the growth conditions. As shown in Fig.~\ref{fig:1}(d), the cross section of the fabricated cavity usually becomes hexagonal, since the crystal growth is strongly affected by the crystal plane of the seed rod. Since the crystal structure of the sapphire is trigonal (hexagonal), the grown cavity is also hexagonal. Although polygonal WGM cavities have advantages in terms of controlled coupling,\cite{kato} it is difficult to obtain an ultra-high $Q$,\cite{ref17, ref18} and a circular shape is often preferred. Hence, to obtain a cavity with a circular cross section, we start the crystal growth after pre-heating the seed rod. Before beginning cavity fabrication, we heated the seed rod with a $\mathrm{CO_2}$ laser at a power of 6~W. We believe this procedure enables the rapid heating and annealing of the seed rod and adds some randomness to the crystal structure. Indeed, we obtained a cavity with a circular cross section as shown in Fig.~\ref{fig:1}(e).

We now have methods for controlling the longitudinal and cross-sectional shape of the grown crystal, which makes it possible to fabricate a bulge-shaped crystalline WGM cavity with LHPG. The crystal quality of the fabricated WGM cavity may be degraded due to the pre-heating procedure. Hence, in the next paragraphs we show our optical measurement result and the x-ray diffraction (XRD) spectra and discuss the tradeoffs between the $Q$, cavity shape and crystal quality.

\begin{figure}[htbp]
\centering
\includegraphics[width=2.8in]{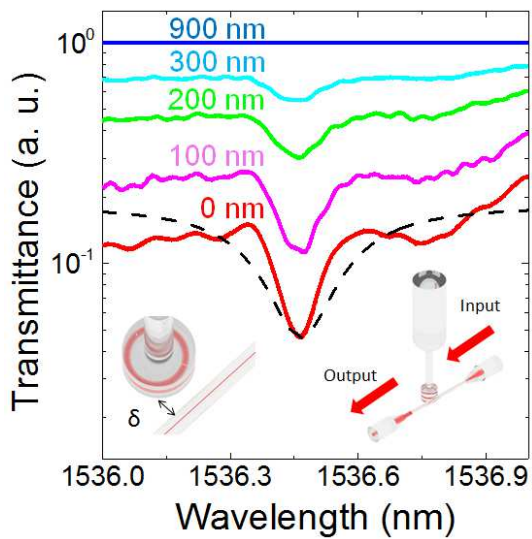}
 \caption{(Color online) (a) Transmission spectrum of the circular cavity shown in Fig.~\ref{fig:1}(e) measured with a conventional tapered fiber setup. We changed the gap distance $\delta$ between the cavity surface and the tapered fiber and measured the transmittance spectrum. $\delta$s are shown in the panel. The dotted line is the Lorenz fit for the spectrum when the tapered fiber is in contact with the WGM microcavity ($\delta=0$~nm). The inset shows the experimental setup.}
 \label{fig:2}
\end{figure}

First we show our optical measurement. We used a conventional tapered fiber setup, with which we used a tapered fiber with a diameter of about $1~\mathrm{\mu m}$ fabricated from a single mode fiber. We controlled the input polarization to excite the TM-WGM mode, and measured the transmittance spectrum by using a wavelength tunable laser (Santec TLS-510). When we measured the transmittance spectrum for a hexagonal cavity (Fig.~\ref{fig:1}(d)), we obtained a $Q$ of $8.5\times 10^3$. On the other hand, we obtained a narrower resonance as shown in Fig.~\ref{fig:2} for a circular cavity (Fig.~\ref{fig:1}(e)). To confirm that we were indeed measuring the resonance of the cavity, we changed the gap distance $\delta$ between the surface of the cavity and the tapered fiber. We observed a larger dip for a smaller distance, which confirms our measurement. By fitting the transmittance spectrum for $\delta=0$~nm with a Lorentz function, we obtained $Q=1.6\times 10^{4}$. A circular cavity exhibits a higher $Q$ thus revealing the importance of cross-section control during crystal growth. The influence of the cavity shape will be discussed later.

\begin{figure}[htbp]
\centering
\includegraphics[width=2.8in]{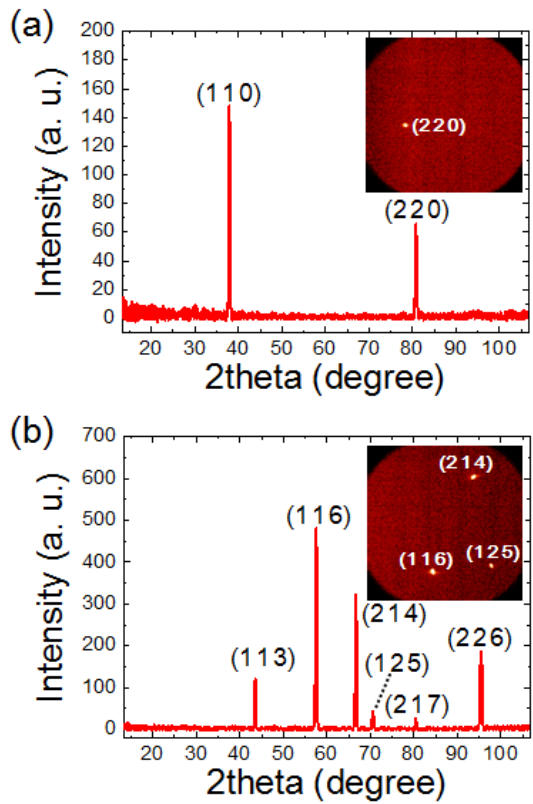}
\caption{(Color online) (a) XRD spectrum of the circular cavity shown in Fig.~\ref{fig:1}(d) measured using $2\theta$-$\theta$ method. The inset is the x-ray diffraction image. (b) As (a) for the cavity shown in Fig.~\ref{fig:1}(e). The corresponding crystal plane orientations of $\mathrm{Al_2 O_3}$ are shown in the panel.}
\label{fig:3}
\end{figure}

Next, we conducted XRD spectrum measurements to analyze the crystal quality of the fabricated devices. We used the $2\theta$-$\theta$ method, which can measure the X-ray diffraction a few $\mathrm{\mu m}$ from the surface, because we are interested in the crystal quality where the WGM exists. Figure~\ref{fig:3}(a) and (b) show the result for hexagonal and circular cavities, respectively. We observed only diffraction peaks that originated from $\mathrm{Al_2 O_3}$. On the other hand, we observed high-order crystal planes in Fig.~\ref{fig:3}(b), which should not appear in our XRD measurement angle if the c-axis was maintained. This indicates the presence of some randomness. However this randomness is limited and the samples retain their crystal structure rather than becoming amorphous, as shown by the bright spots in the 2D diffraction image in the inset of Fig.~\ref{fig:3}(b). In brief, there is a trade-off between the crystalline and the shape; however, the quality of the crystal was not significantly degraded, even for circular cavity fabrication. The experimental optical measurement and XRD results suggest that the cross sectional shape of the cavity is an important factor in determining the $Q$.

Finally, with this in mind, we performed a numerical analysis based on 2D finite-difference time-domain (FDTD) calculations to study the effect of the cross sectional shape of the WGM cavity.

\begin{figure}[htbp]
\centering
\includegraphics[width=2.8in]{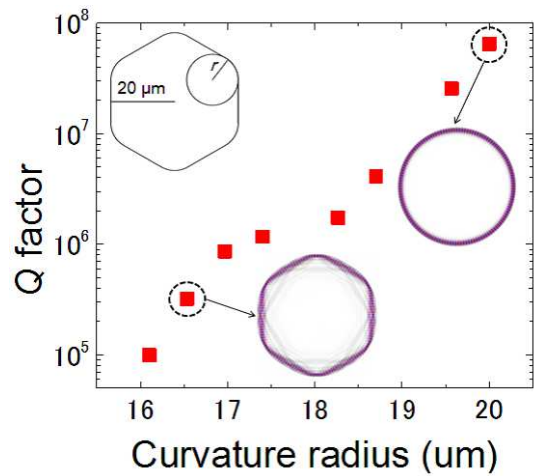}
\caption{(Color online) Calculated $Q$-factors with respect to the edge curvature radius $r$ of the hexagonal WGM cavities. The inset in the upper left is the structural model used for the calculation. The cavity radius is $20~\mathrm{\mu m}$ and the refractive index is 2.4. The profiles of the perturbed-WGM for $r=0~\mathrm{\mu m}$ and $16.5~\mathrm{\mu m}$ are shown in the panel.}
\label{fig:4}
\end{figure}
Figure~\ref{fig:4} shows that the $Q$ of the WGM mode is higher when the cross section of the cavity becomes close to a circle. Note that the WGM mode we found in Fig.~\ref{fig:4} is different from that studied in ZnO hexagonal cavities\cite{ref17, ref18}. In contrast to the light trace of the known quasi-WGM, which reflects at the side of the cavity, the light of this perturbed-WGM propagates close to the surface at the corner of the polygon and far from the surface at the side of the cavity\cite{kato}. Since this mode originates from a regular WGM that is available in a circular cavity, it has a much higher $Q$ than a quasi-WGM. Analysis shows that $r$ is the critical parameter for obtaining high $Q$, and indeed, we controlled this parameter by pre-heating the seed rod.

In summary, we demonstrated the fabrication of a crystalline resonator using the LHPG technique. We obtained a WGM cavity made of sapphire with a $Q$ of $1.6\times 10^4$. A high-$Q$ cavity was obtained because we successfully fabricated a WGM cavity with a circular cross section. XRD measurement revealed that the overall crystalline structure was maintained.

This work was partly supported by Keio University's Program for the Advancement of Next Generation Research Projects.

%%%%%%%%%%%%%%%%%%%%%%% References %%%%%%%%%%%%%%%%%%%%%%%%%

\end{document}